\documentclass[12pt,preprint]{aastex}
\usepackage{psfig}
\def\HI {H\kern0.1em{\sc i}} 
\def\deg{$^{\circ}$}

\slugcomment{Accepted by Astrophysical Journal}

\begin{document}
\title{~~\\ Polarimetry of Compact Symmetric Objects~~\\ }
\shorttitle{CSO Polarization}
\shortauthors{Gugliucci et al.}
\author{N.E. Gugliucci\altaffilmark{1,3},
  G.B. Taylor\altaffilmark{2,3}, A.B. Peck\altaffilmark{4}, and
  M. Giroletti\altaffilmark{5,6}}
%\affil{}
\email{neg9j@virginia.edu; gbtaylor@unm.edu; apeck@cfa.harvard.edu; giroletti@ira.inaf.it}
\altaffiltext{1}{Department of Astronomy, P.O. Box 400325, University of Virginia,
  Charlottesville, VA 22904; neg9j@virginia.edu}
\altaffiltext{2}{Department of Physics and Astronomy, University of
  New Mexico, Albuquerque, NM 87131; gbtaylor@unm.edu}
\altaffiltext{3}{National Radio Astronomy Observatory, P.O. Box O,
  Socorro, NM 87801}
\altaffiltext{4}{Harvard-Smithsonian Center for Astrophysics, SAO/SMA
  Project, 645 North A'ohoku Place, Hilo, HI 96720; apeck@cfa.harvard.edu}
\altaffiltext{5}{INAF Istituto di Radioastronomia, via Gobetti 101,
40129 Bologna, Italy; giroletti@ira.inaf.it}
\altaffiltext{6}{Dipartimento di Astronomia dell'Universit\`a di Bologna, via Ranzani 1, 40127 Bologna,
Italy}
%\Received{}
%\accepted{}
%\journalid{}{}
%\articleid{}{}

%\slugcomment{DRAFT}

\begin{abstract}

We present multi-frequency VLBA observations of two polarized Compact
Symmetric Objects (CSOs), J0000$+$4054 and J1826$+$1831, and a
polarized CSO candidate, J1915$+$6548.  Using the wavelength-squared
dependence of Faraday rotation, we obtained rotation measures (RMs) of
$-$180$\pm$10~rad~m$^{-2}$ and 1540$\pm$7~rad~m$^{-2}$ for the latter
two sources.  These are lower than what is expected of CSOs (several
1000~rad~m$^{-2}$) and, depending on the path length of the Faraday
screens, require magnetic fields from 0.03 to 6~$\mu$G.  These CSOs
may be more heavily affected by Doppler boosting than their
unpolarized counterparts, suggesting that a jet-axis orientation more
inclined towards the line of sight is necessary to detect any
polarization.  This allows for low RMs if the polarized components are
oriented away from the depolarizing circumnuclear torus.  These
observations also add a fourth epoch to the proper motion studies of
J0000$+$4054 and J1826$+$1831, constraining their kinematic age
estimates to $>$610~yrs and 2600$\pm$490~yrs, respectively.  The
morphology, spectrum, and component motions of J1915$+$6548 are
discussed in light of its new classification as a CSO candidate, and
its angle to the line of sight ($\sim$50\deg) is determined from
relativistic beaming arguments.

\end{abstract}


\section{Introduction}

Compact symmetric objects (CSOs) are a class of active galactic nuclei
(AGN) that have significant visible jet or hotspot activity on either
side of the central engine \citep{con94, wil94}.  These are typically
$<$~1~kpc in size because they are young objects
\citep[$\leq$3000~yr;][]{pol03, gug05}.  In terms of the unified
scheme of AGN \citep{ant93}, there is little Doppler boosting of the
jets and hotspots due to the orientation of the source with respect to
the line of sight.  This orientation also allows for studies of the
hypothesized dust and gas torus that surrounds the central engine and
lies perpendicular to the jet axis.  The actual angle that the line of
sight makes with the torus determines what features can be seen.
Evidence for this torus comes from detections of broad HI absorption
lines \citep{tay99, pec00b, pih03, gup06} and free-free absorption
\citep{pec99} towards CSOs.  H$_2$O megamasers also provide
clues to circumnuclear tori in nearby AGN \citep[][and references
therein]{lo06}.

This torus, and magnetic fields in the shocked, photoionized ISM
around the lobes, may also give rise to the large Faraday screens that
depolarize emission from most CSOs \citep{bic97}.  The first detection
of significant polarization in two CSOs in the COINS sample
\citep[CSOs Observed In the Northern Sky;][]{pec00a}, J0000$+$4054
(2.1\%) and J1826$+$1831 (8.8\%), occured on the side with the more
prominent jet or hotspot \citep[][hereafter GTPG]{gug05}.  The
opposite hotspot showed no polarization down to a 0.3~mJy limit.  Any
polarization that is detected should have high rotation measures
\citep[several 1000s rad m$^{-2}$;][and references therein]{bic97}.
In this paper, we attempt to calculate the Faraday rotation measures
in these polarized components.

Kinematic ages can be obtained by measuring the separation speed
between hotspots over time, or the proper motion of a hotspot or jet
component with respect to the core \citep[e.g.][]{pol03, nag06}.  In
GTPG, we confirmed ages for three CSOs between 130$\pm$47 and
3000$\pm$1490 yr, with another source appearing to have an age of
20$\pm$4 yr.  The overall CSO age distribution seems to be
disproportionately stacked towards the younger ages.  Possible
explanations are: there is a selection effect against older CSOs,
the jet activity tends to die off after a certain period of time,
or CSOs have periods of deactivation and reactivation
\citep{tin03}.  Those that survive may evolve into Fanaroff-Riley II
galaxies such as Cygnus A \citep{fan74, rea96b, pol03}.  In this
paper, we refine the kinematic age estimates of J0000$+$4054 and
J1826$+$1831 that were first calculated in GTPG.  Also included are
observations of a new CSO candidate from the Second Caltech-Jodrell
Bank Survey, J1915$+$6548, which also shows hotspot separation over
three epochs and polarization in its more prominent hotspot.

Throughout this discussion, we assume H$_{0}$=71 km s$^{-1}$
Mpc$^{-1}$, $\Omega_M$ = 0.27, and $\Omega_{\Lambda}$= 0.73.  Linear
sizes and velocities for sources with known redshifts have been
calculated using E.L. Wright's cosmology calculator
\footnote{http://www.astro.ucla.edu/$\sim$wright/CosmoCalc.html}.

\section{Observations and Analysis}

Observations were centered on 4.8~GHz and 8.4~GHz on 12 February 2005 for
J0000$+$4054 and on 18 February 2005 for J1826$+$1831 using the
VLBA\footnote {The National Radio Astronomy Observatory is
operated by Associated Universities, Inc., under cooperative
agreement with the National Science Foundation.}. Observations of
J1915$+$6548 were centered on 4.8~GHz, 8.4~GHz, 15.1~GHz, and 22.2~GHz
with the VLBA on 11 November 2004.  Each frequency was
separated into four IFs, and these IFs were paired such that the higher
two frequencies were averaged during imaging as were the lower two
frequencies, except at 22.2~GHz where all four IFs were averaged.
Therefore, the frequencies used for the RMs in these
observations were 4.6~GHz, 5.0~GHz, 8.2~GHz, 8.5~GHz, 14.9~GHz,
15.3~GHz, and 22.2~GHz.  Observational parameters are presented in
Table 1.

Amplitude calibration of the data was derived from system temperatures
and antenna gains.  Fringe-fitting was performed with the AIPS task
FRING on the calibrators OQ~208 and 3C~84.  D-term solutions were
determined with the AIPS task LPCAL and the same calibrators OQ~208
and 3C~84.  Absolute electric vector position angle (EVPA) calibration
was determined for J0000$+$4054 using the EVPAs of J1310$+$322 and for
J1826$+$1831 and J1915$+$6548 using the EVPAs of BL~Lac listed in the
VLA Monitoring
Program\footnote{http://www.vla.nrao.edu/astro/calib/polar/}
\citep{tmy00}.  Note that the EVPAs were corrected for each of the
four IFs separately.

\section{Results}

Faraday rotation was first noted by Michael Faraday when he passed
polarized light through a refractive medium in the presence of a
magnetic field \citep{far33}.  The intrinsic polarization angle,
$\chi_0$, is observed as $\chi$ such that
\begin{equation}
\chi = \chi_0 + RM\lambda^2
\end{equation}
where $\lambda$ is the observed wavelength.  The rotation measure, RM,
is related to the electron density, $n_{\rm e}$, the net line of sight
magnetic field in the environment, $B_\|$, and the path length, $dl$,
through the plasma, by the equation
\begin{equation}
RM = 812\int n_{\rm e} B_\| dl \quad \mbox{rad m$^{-2}$}
\end{equation}
where units are in cm$^{-3}$, mG, and parsecs.  A reasonable $n_{\rm
e}$ for radio galaxies is 10$^3$~cm$^{-3}$ as estimated by
\cite{zav03}.  Our lower limit for the path length is 0.3~pc, the
approximate size of clumps within the Faraday screen in M87
\citep{zav02}.  Since the Faraday screen for an AGN is now considered
to come from interactions of the jet with ambient material
\citep{zav04}, a good upper limit for the path length is the jet
radius, or 10~pc.  Field strengths calculated with these parameters
can be compared to the strength of a magnetic field in pressure
balance with a thermal gas of the same $n_{\rm e}$ and a temperature
of 10$^4$~K using
\begin{equation}
{B^2 \over {8\pi}} = {n_{\rm e} k T \mbox{.}}
\end{equation}

The 8.4~GHz images of J0000$+$4054 and J1826$+$1831 at full resolution
are shown in Figures 1 and 2 with sticks representing polarization
vectors.  The 8.4~GHz images were then tapered to match the resolution
of the 4.8~GHz images.  The same circular beam was then applied to
both images and spectral index maps were created.  These are overlayed
with the 4.8~GHz images in Figures 1 and 2.  A plot for the rotation
measure of the polarized component of J1826$+$1831 is shown in Figure
3.  Source parameters for these CSOs and J1915$+$6548 are in Table 1.

In GTPG, we attempted to obtain relative proper motions for CSOs in
the COINS sample in order to calculate a kinematic age for each
source.  This assumes that the separation speed of the hotspots is
uniform.  This method is independent of the angle that the source
makes with respect to the line of sight and of the distance to the
source.  This method also produces results if the hotspot or outer jet
component can be seen moving away from the core.  Using 8.4~GHz data
from three epochs spread across a five year period, we calculated the
kinematic ages of three CSOs with reasonable certainty, while the rest
provided lower limits.  These new observations provided us with a
fourth epoch for J0000$+$4054 and J1826$+$1831 at 8.4~GHz, extending
the time baseline to seven years.  This provides better estimates and
limits of the ages of these two CSOs.

Models for J0000$+$4054 and J1826$+$1831 were fit to the visibility
data for the March 2000 epoch, since this was close to the middle of
the time baseline and had a high dynamic range.  These models were
comprised of elliptical Gaussians.  Visibility data from each epoch
were fit to a model with the same size and shape ellipses so that only
the positions and fluxes of these were allowed to vary with time.
Uncertainties in position for each component were calculated from the
signal-to-noise ratios and the synthesized beam.  The positions of
these components with respect to a reference component were fit with a
weighted least squares line where the slope of the line provides the
relative speed.  Motions are considered significant if they are at
least 3$\sigma$ above the errors and if the reduced chi-squared is
nearly 1.  Errors are dependent on the image noise and the individual
flux of each component.  If large errors are present, then the motions
are considered upper limits, so that they give rise to ages that are
lower limits.  Table 3 includes the modelfit parameters, and Figure 4
gives plots of the proper motion of two components in J1826$+$1831.

Total intensity images of J1915$+$6548, the new CSO candidate, are
presented in Figure 5 with sticks representing polarization vectors
overlayed.  The integrated total power spectrum, as well as that of
component A alone, is shown in Figure 6.  Since the brightest hotspot
was polarized at 8.4~GHz, 15.1~GHz, and 22.2~GHz, a rotation measure
plot is given in Figure 7.  Figure 8 presents the proper motions of
hotspot separation in this source at 4.8~GHz over an 11 year timespan.
Table 4 gives hotspot brightnesses for the polarized sources and the
unpolarized sources of the COINS sample for comparison in $\S$4.
Plots of $\beta$ vs. $\theta$ for the separation speeds and
orientations of J1826$+$1831 and J1915$+$6548 are presented in
Figure~9.

\subsection{\bf J0000$+$4054}  This CSO was identified as such in
\cite{dal02} and GTPG, and polarization was detected at 8.4~GHz.  It
is associated with a galaxy of magnitude 21.4 \citep{sti96}.  The core
has still not been positively identified from the spectral index map
(see Fig. 1).  From their compact morphologies, either B1 or B2 could
be the core, but their spectral indices are rather steep ($\alpha
\approx -$0.6; $S_{\nu} \propto \nu^{\alpha}$).  However, there
appears to be a flattening ($\alpha \approx -$0.4) between components
B1 and B2, so the core may be located there (see inset Fig. 1.b).  In
\cite{dal02}, the combined flux of B1 and B2 (component Ce in that
paper) is 71~mJy at 1.6~GHz.  When compared with the fluxes of B1 and
B2 here at 8.4~GHz, this yields a spectral index for that region of
$\alpha \approx-$0.25.  High dynamic range imaging at 15~GHz should
provide a positive identification.

Polarization was detected in the southern hotspot at 8.4~GHz with an
intensity of 2.1~mJy (see Fig. 1).  This is 1.2\% of the intensity of
the hotspot. There is no detectable polarization at 4.8~GHz down to
the 3$\sigma$ level ($\sim$ 0.2~mJy, or 0.06\%).  The 4.8~GHz image
also shows a strange morphology in the southern lobe.  There appears
to be a hole, or depression, in the emission below the bright hotspot.
If this is not an artifact in the data, it resembles the region in the
eastern lobe of 4C~31.04 by \cite{gir03}, where it is speculated that
such a hole could be created by a dense molecular gas that is
impenetrable to radio emission or to the plasma itself.  However, the
morphology may also be the effect of edge-brightening at component C
if the plasma is running into denser material.  This material may also
be deflecting the jet to the east, as seen in the 8.4~GHz image.
Polarization was only detected at 8.4~GHz, so the RM could not be
determined and the true orientation of the magnetic field is not
known.  If the RM was known to be as low as in J1826+1831
($-$180~rad~m$^{-2}$, see $\S$3.2), for example, then there would be
little rotation from the original polarization angle to 8.4~GHz, so
the magnetic field would lie roughly perpendicular to the electric
vectors in Figure 1a.  Then, the detectable polarization in this
region would be due to a compression of the ambient magnetic field at
a shock front where the jet is colliding with the denser gas.  Future
observations at 15 GHz should allow us to determine the rotation
measure for this region.

The fourth epoch of 8.4~GHz data in the proper motion study of this
object did not provide a good fit for the separation speed of the
hotspots.  However, there is an upper limit of 0.066~mas~yr$^{-1}$ of
component A away from component C.  This speed limit and a hotspot
distance of 40.33~mas provides a lower limit for the kinematic age of
610~yrs.  This is higher than was previously estimated (280~yrs in
GTPG).  Although the redshift is not known, a reasonable estimate of
$z\approx$~0.5 can be made for the typical redshift of a CSO host
\citep{aug06}.  If this is the case, the projected distance between
components A and C is 242~pc and A is moving with speed $v<$ 2.0$c$.

\subsection{\bf J1826$+$1831}  This is the most significantly
polarized of these three objects.  Component C, most likely a jet
component, has a polarized intensity of 2.3~mJy at 8.4~GHz, or 8.5\%
of its flux (see Fig. 2).  This component has a polarized intensity of
1.4~mJy at 4.8~GHz, making it 3.7\% polarized.  A least-squares fit
for the rotation measure of $-$180$\pm$10~rad~m$^{-2}$ at the peak of
component C is presented in Figure 3.  This is not an uncommon
rotation measure for a typical quasar jet \citep{zav04}.  The pair of
angles at 8.2 and 8.5~GHz suggest a higher rotation measure, but this
requires for a number of 180 degree turns to be put in between 4.6 and
5.0~GHz and between 5.0~GHz and 8.2~GHz.  These turns can be
introduced because the polarization vectors give an orientation, but
not a direction, of the electric field such that $\chi$ and $\chi
\pm$180 are indistinguishable.  Depending on the number of turns used
in this data, rotation measures as high as 6000~rad~m$^{-2}$ are
plausible.  One must take caution with this, however, since any number
of turns can be introduced to provide a false good fit.  Therefore,
$-$180$\pm$10~rad~m$^{-2}$ is a conservative estimate.  Measurement of
polarization at another frequency may help in determining the correct
rotation measure.  It may also be true, however, that we only see
polarized emission from this component because it is free of the high
rotation measures that may be present closer to the core.

If a path length of 0.3~pc is assumed with n$_e =$ 10$^3$~cm$^{-3}$,
the magnetic field strength is 0.7~$\mu$G, using Eqn.~2.  With the
upper limit of the path length, 10~pc, the field strength is only
0.03~$\mu$G.  Both of these are much smaller than the strength of a
magnetic field that is in pressure balance with a 10$^4$~K gas of the
same electron density, $\sim$200~$\mu$G, using Eqn.~3.

A spectral index map was made between 4.8 and 8.4~GHz which is
presented in Figure 2.  This shows the core, B, to be a flat spectrum
component ($\alpha \approx$ 0.6), while the rest of the jet and
counterjet are steep spectrum ($\alpha \approx$ $-$0.6 to $-$1.7).

The proper motions of components C and D with respect to B were
measured using the four epochs of 8.4~GHz data available to us.  D was
found to be moving at 0.015$\pm$0.003~mas~yr$^{-1}$.  With a projected
separation between B and D of 41.87 mas, this corresponds to a
kinematic age for the CSO of 2600$\pm$490~yr.  This refines our
earlier estimate of 3000$\pm$1490~yr.  The jet component C is moving
along at a faster rate of 0.032$\pm$0.003~mas~yr$^{-1}$ (see Fig. 4).
The modelfitting of components works best for bright, compact
components, of which component A is neither, so no proper motions can
be fit.  If we assume again that $z \approx$ 0.5, D is moving away from B
at $\sim$0.45$c$ and is 255~pc away.  C is moving away from B at
$\sim$0.95$c$.  However, this source has no detected optical
counterpart, so its redshift could be much higher.

\subsection{\bf J1915$+$6548} This new CSO candidate was first listed
in a survey with the 300-ft Green Bank Telescope at 4.8~GHz by
\cite{bec91} It was observed in the Second Caltech-Jodrell Bank Survey
(CJ2) which presented a spectrum of total intensity in \cite{hen95}.
Its host is a Seyfert~1 galaxy with a magnitude of 18.2 and redshift
of $z =$ 0.486 \citep{hen97}.  New observations were taken with the
VLBA to investigate the non-linear morphology in the CJ2 image.

We identify Component A as a steep spectrum hotspot ($\alpha \approx
-$0.9; See Fig. 6) that dominates the flux of this object at all four
frequencies.  Components B and C are steep spectrum ($\alpha \approx$
$-$1.2 and $-$0.9, respectively), and we propose that they are also
part of the eastern lobe.  Component D is very steep spectrum ($\alpha
\approx$ $-$1.6), and, based on morphology, we propose it to be the
counter-hotspot to A.  There is no evidence of a compact, flat or
inverted spectrum core, but there are a growing number of CSOs that
have two steep spectrum hotspots but no visible core (e.g. J0620+2102,
J1111+1955; GTPG).  There is also a trail of extended emission between
A and D at 4.8 and 8.4~GHz that hints at a jet-like structure, as well
as emission further east of A in the 4.8~GHz image which we interpret
as a sign of earlier activity (see Fig. 5).  Although the morphology
is similar to that of a CSO, the hotspot intensity ratio at 4.8~GHz is
12:1 and is 26:1 at 8.4~GHz.  These fall short of the 10:1 criteria
set for CSOs in the COINS sample \citep{pec00a}.

To determine the synchrotron age of the source, we present the total
intensity spectrum of J1915$+$6548 in Figure 6 using total intensities
from this paper and data points from the NASA/IPAC Extragalactic
Database (NED)\footnote {NED is operated by the Jet Propulsion
Laboratory, Caltech, under contract with NASA.}.  Using the break
frequency, 17~GHz, the minimum energy magnetic field of the plasma can
be calculated as in \cite{mil80} using
\begin{equation}
B_{me} = 1.4 \times10^{-4}(1+z)^{1.1}\nu^{0.22}_0{\left({F_0}\over
  {\theta_x\theta_ys}\right)}^{2/7}
\end{equation}
where B is in gauss, $\nu_0$ is in GHz, F$_0$ is in Jy, $\theta_x$ and
$\theta_y$ are the dimensions of an elliptical component in
arcseconds, and $s$ is the path length in kpc.  With this, the
synchrotron age of the source can be calculated as in \cite{mur99}
with
\begin{equation}
\tau_{syn} = {{1610 B^{0.5}}\over{[B^2+B^2_{CMB}][\nu_{br}(1+z)]^{1/2}}}
\end{equation}
with B in $\mu$G, $\nu_{br}$ in GHz, and $B_{CMB} =$ 3.25(1$+z$)$^2$.
At 4.8~GHz, the eastern hotspot is 0.291~Jy and can be approximated by
an ellipse 0.01~mas by 0.006~mas.  We assume the path length to be
equivalent to 0.01~mas, or 0.07~kpc at a redshift of 0.486.  This
yields a magnetic field of $\sim$7~mG and a synchrotron age of
$\sim$540~yrs.

Component A is polarized at 8.4~GHz (3.0~mJy or 1.6\%), 15.1~GHz
(2.3~mJy or 2.8\%), and 22.2~GHz (1.8~mJy or 4.0\%).  The polarized
flux is too low to be reliable at 4.8~GHz (1.3~mJy or 0.5\%).  A plot
of the rotation measure is given in Figure 7.  The rotation measure of
1540$\pm$7~rad~m$^{-2}$ comes from a direct measurement of the EVPAs
at the center of the polarized component without any introduced 180
degree wraps.  This is more consistent with the high RMs expected from
a CSO.  With a path length of 0.3~pc and the same electron density as
above, this corresponds to a magnetic field strength of 6~$\mu$G.  At
the upper limit for path length of 10~pc, the magnetic field strength
is 0.2~$\mu$G.  These are still much lower than what is expected in
the central regions of a radio galaxy (200~$\mu$G, see $\S$3.2).

The visibility data at 4.8~GHz was modelfit at this epoch and at two
other epochs, 1993.444 and 1995.691, the first of which was presented
in \cite{hen95}.  Proper motions were fit with a least squares line
with a slope of 0.035$\pm$0.004~mas~yr$^{-1}$.  With a redshift of $z
=$0.486, this apparent speed corresponds to 1.02$\pm$0.11~$c$.
Although this is higher than typical hotspot separation speeds (GTPG),
it yields a good fit (see Fig. 8) and a kinematic age of
940$\pm$110~yr.

\section{Discussion}

\subsection{CSO Orientations}

Depolarization in CSOs is consistent with the \cite{bic97} model.
Variations in rotation measure across the lobe can cause this
depolarization, and this may be due to magnetic field reversals in the
post-shock ISM.  The obscuring torus may also play a role in
depolarization, and this would be dependent on torus scale height,
opening angle, and orientation.  We rule out depolarization by a thin
disk of material, with a scale height much less than the radio source
size, since we frequently see that both sides are depolarized, whereas
a thin disk would only cover one side.  If a Faraday screen is to
produce the observed polarization asymmetry in these sources, then it
requires a scale height of about half the total source size.  More
detections of polarization in CSOs are needed to constrain its
geometry.

The existence of polarized emission in these three sources sets them
apart from the rest of the COINS sample.  However, this is not the
only difference.  In GTPG, we noted that J0000$+$4054 and J1826$+$1831
appeared to have greater hotspot intensity ratios than their
unpolarized counterparts.  Since the polarization is detected on the
side of the more prominent hotspot, this can be attributed to a
shorter path length through the circumnuclear torus and, consequently,
a lower Faraday depth.  Table 4 lists the brightnesses of each hotspot
and the ratio of the dimmer one to the brighter one for unpolarized
and polarized CSOs.  There is a bimodal distribution in that the
polarized sources presented in this paper have ratios $\leq$0.1 and
the unpolarized COINS have ratios $\geq$0.3.  Although the sample size
for polarized CSOs is small, this is consistent with the observed
polarization being due to an orientation effect.  As the angle between
the jet axis and line of sight decreases, the forward hotspot may be
more Doppler boosted and the receding hotspot Doppler dimmed, assuming
mildly relativistic bulk motions.  So, the leading hotspots in the
three polarized sources are oriented such that they are relatively
free from Faraday depolarization by the torus that surrounds the
central region.  However, to explain the depolarization over many
viewing angles in the other CSOs, a large torus is needed to extend
over hundreds of parsecs of jets and lobe.  HI absorption has been
seen at this distance from the central engine
\citep[e.g. 1946+708][]{pec99}.

In order to quantify depolarization in CSOs, we look at Faraday beam
depolarization.  We can calculate the necessary rotation measure
gradient across the synthesized beam in order to get a rotation of one
radian.  For the unpolarized CSOs in the COINS sample, with an average
beamwidth of 1.5~mas at 8.4~GHz, the RM gradient need only be
$\sim$500~rad~m$^{-2}$~mas$^{-1}$.  So, the RM of J1915$+$6548 may be
reasonable for a CSO, whereas the low rotation measure of J1826$+$1831
is still puzzling.  An RM of 500~rad~m$^{-2}$ corresponds to a
magnetic field strength of 2~$\mu$G with a path length of 0.3~pc and
0.06~$\mu$G with 10~pc.  These are still much lower than the strength
of a magnetic field in pressure balance with a 10$^4$~K gas.

We use the assumption that relativistic beaming is in effect to quantify
the orientations of these sources.  We can match up what appear to be
jet components from the approaching and receding sides of the core and
compare their fluxes ($S_a$ and $S_r$) as
\begin{equation}
{S_a \over S_r} = \left({1+\beta\cos{\theta} \over 1-\beta\cos{\theta}}\right)^{n-\alpha}
\end{equation}
where $\beta$ is the space velocity, $\theta$ is the angle to the line
of sight, $\alpha$ is the spectral index, and $n$ is either 2 or 3.
Models for continuous jets are best fit by $n =$ 2 and jets of
discrete components by $n =$ 3.  One can use
\begin{equation}
{{\mu_a} \over {\mu_r}} = {{d_a} \over {d_r}} = {{1 + \beta
\cos{\theta}} \over {1 - \beta \cos{\theta}}}\mbox{,}
\end{equation}
where $\mu$ is the apparent motion and $d$ is the distance from the
core.  However, this requires a well known position for the center of
activity.  One can also use the hotspot separation, $\mu_{sep} =
|\mu_a| + |\mu_r|$, such that
\begin{equation}
{v_{sep}} = {\mu_{sep}D_a(1 + z)} = {{2{\beta}\sin{\theta}} \over {1 -
    \beta^2\cos^2{\theta}}}\mbox{,}
\end{equation}
where $D_a$ is the angular size distance to the source, $z$ is the
redshift, and $v_{sep}$ is the angular separation speed in units of
$c$ \citep{tay97}.  These can provide constraints on $\beta$ and
$\theta$.

We know neither the core location nor the redshift for J0000$+$4054.
So, only Eqn.~6 can be applied, using the fluxes of the hotspots as
determined by the model in Table~3.  A locus of $\beta$ and $\theta$
is plotted in Figure~9a.  Note that a lower limit is plotted when $n
=$ 2 and an upper limit when $n =$ 3.  The errors in the measured
fluxes are negligible compared to this spread.  This shows that,
approximately, $\beta \geq$ 0.3 for all values of $\theta$.

For J1826$+$1831, we used the brightnesses of each hotspot, components
A and D, with Eqn.~6 as well as their distances from the core with
Eqn.~7 to calculate loci of $\beta$ and $\theta$ (Figure~9b).  Note
that these two equations have a similar functional form, so no tight
constraints can be determined.  This gives an upper limit for the angle
to the line of sight of $\sim$75$^{\circ}$.  The true space velocity
will be $\geq$0.3~$c$ for any angle.  This is higher than typical
values found for other CSOs \citep[$\sim$0.1$c$][]{ows98a} and other
radio galaxies \citep[$<$0.1$c$][]{sch95}.  If the value for
$\mu_{sep}$ of component D (0.015~mas~yr$^{-1}$)is used in Eqn.~8 with
a typical CSO redshift ($z\approx$ 0.5), the projected separation
velocity is $\sim$0.4$c$.  This is consistent with our results.
J1826$+$1831 cannot be very close (for example, $z=$ 0.01), for then
the separation velocity would be unusually low and inconsistent with
Figure~9b.

%Although the redshift is not
%known, the angular speed of component C from the core exceeds $c$
%around $z =$~0.5.  Since the optical counterpart is faint and no
%redshift has been obtained to date, this argues for a more distant
%object where superluminal motion, and therefore Doppler boosting,
%becomes a factor.

We used the apparent hotspot separation speed of 1.02$\pm$0.11$c$ for
J1915$+$6548, $z=$0.486, and Eqn.~8 to plot the locus for $\beta$ and
$\theta$ in Figure~9c along with the loci for the two brightness
models with Eqn.~6.  These two sets of curves intersect when
$\beta \approx$ 0.55 and $\theta \approx$ 50\deg.  Such a high space
velocity is uncharacteristic of lobes.  However, it is possible that
the hotspot being measured is a transient feature, and therefore
moving more quickly, whereas the lobe as a whole is moving at a much
slower speed.  This can occur if the jet that is feeding the hotspot
has changed its orientation slightly and is drilling out a new part of
the lobe \citep{sch95}.  Alternatively, we may be measuring the
separation of jet components rather than hotspots as there may be no
visible hotspots.  This example shows how the orientation angle of a
CSO can be measured if mild relativistic beaming is a factor.

%The above analysis cannot be done for J0000$+$4054 because it does not
%have a known redshift or a definite core location.  If both the
%redshift and core are known, Eqns.~6 and 7 can provide loci for
%$\beta$ and $\theta$ that intersect and provide constraints on the
%source geometry and kinematics, provided that the source morphology is
%dominated by Doppler boosting effect.  For CSOs in the dense nuclear
%environment, local interactions may frequently dominate the
%appearance.

\subsection{CSO Environments}

An alternate theory for the small sizes of CSOs is that they are old
sources frustrated by a dense medium \citep{car94}.  This dense medium
may in fact be asymmetric, possibly as a result of a galaxy merger or
interaction \citep{car98}.  In an extreme case, the ages of these
sources would be comparable to the ages of larger, classic radio
doubles.  \cite{rea96a} pointed out that if this were the case, CSOs
would have more spherical morphologies.  Also, hotspot separation
speeds in \cite{pol03} and GTPG show that the hotspots are still
moving too quickly to be confined by a dense medium and be older than
a few thousand years.  However, this does not rule out high densities
or asymmetries in the medium around CSOs as evidence by \cite{ori06}.
\cite{bic97} propose that CSOs are frustrated but not confined by the
interstellar medium (ISM).  Their model assumes a dentist-drill
explanation for jet-lobe interactions and predicts low polarization.
This is because the ionized gas surrounding the lobes in this model
produces large variations in Faraday rotation measures across the
source, thus depolarizing the emission.  Interstellar magnetic fields
can play an integral role if there are a large number of magnetic
field reversals across the source which would produce a varied RM
structure in our maps.  Since the polarization is spatially isolated
in these CSOs, no such maps can be made.

Asymmetries in the CSO environment may cause one hotspot to be
considerably more polarized in these few sources by interactions with
a dense ISM.  If the EVPAs for J1915+6548 are extrapolated back to
zero wavelength, the observed angle of the electric field is
$-$88$^{\circ}$.  So, the orientation of the magnetic field in the
image would be nearly north-south, which is approximately
perpendicular to the source orientation.  This may be an indication of
a collision between the bright hotspot and a dense medium that orders
the magnetic field along the axis of compression.  This is also
observed in J1826$+$1831 where the EVPA at zero wavelength is
75$^{\circ}$, so the magnetic field is oriented at $-$25$^{\circ}$.
This is also nearly perpendicular to the source orientation, but this
effect is for the bright jet component, not the hotspot.  Components
may also be brighter (by Doppler boosting) if they are moving more
quickly through a less dense ISM.  However, the indistinguishable
motions of the polarized component in J0000$+$4054 seems to refute
that.

This asymmetry in environments does not, however, explain the bimodal
distribution in Table~4, that is, the fact that the ratios of the
dimmer hotspots to the brighter hotspots of each source are much lower
for polarized CSOs than for unpolarized CSOs.  This is more naturally
explained by Doppler boosting.  Also, the relativistic hotspot speeds
of J1915+6548 (see Fig.~9) could not exist in a dense environment.
Finally, we do not see rotation measures in the several thousands as
predicted by the \cite{bic97} model.

\section{Conclusions}

In this study we have investigated the polarization properties of two
CSOs, J0000$+$4054 and J1826$+$1831, and a CSO candidate J1915$+$6548.
The Faraday rotation measures that were observed for J1826$+$1831 and
J1915$+$6548 are lower than what was previously expected for CSOs,
$-$180$\pm$10~rad~m$^{-2}$ and 1540$\pm$7~rad~m$^{-2}$, compared to
predicted values of several thousand ~rad~m$^{-2}$.  These imply low
external magnetic field strengths, 0.03 to 6~$\mu$G, depending on what
estimates for electron density and path length are used.  However,
only about 500~rad~m$^{-2}$~mas$^{-1}$ is needed for beam
depolarization to be a major factor for most unpolarized CSOs.

Since these polarized sources have significantly more asymmetric
hotspot brightnesses than their unpolarized counterparts, with ratios
of $\leq$0.1 as opposed to $\geq$0.3, it is plausible that Doppler
boosting is in effect for the brighter hotspots.  Then, their jet axis
orientations are closer to the line of sight than for other CSOs such
that the polarized hotspot is free of obscuration from the
circumnuclear torus.  We put contraints on $\beta$ and $\theta$ for
all three sources (Figure~9) such that for J0000$+$4054 and
J1826$+$1831, $\beta \geq$ 0.3.  More information, specifically the
redshift, must be known in order to make tighter constraints.  For
J1915$+$6548, $\beta \approx$ 0.5 and $\theta \approx$ 50\deg.

We extended the time baseline for proper motions of the two objects
from the COINS sample from five to seven years.  This provided a
better age estimate for J1826$+$1831 of 2600$\pm$490~yrs and a better
lower limit for the age of J0000$+$4054 at 610~yrs.  Extending the
time baseline for the proper motions of a larger sample of CSOs will
reduce some of the uncertainties and lower limits of their ages so
that the distribution of ages in the sample can be better determined.

J1915$+$6548 was determined to be a CSO candidate because of the
presence of symmetric, steep spectrum hotspots and the lack of a
compact, flat or inverted spectrum core.  If confirmed, this will be
the third CSO with significant polarization.  Although spectral age
arguments and proper motions from an eleven year time baseline suggest
an age between approximately 600 and 1000~yrs, the hotspot separation
speed is most certainly relativistic and Doppler boosting is occuring
in the brighter hotspot.  Our age estimates may be underestimating the
true source age if the brightest hotspot is undergoing renewed
activity from a slowly moving jet or if it is encountering a
particularly dense medium, reaccelerating the electrons and giving a
lower synchrotron age.  If it were the case that the hotspots are
transient features and therefore appear younger and faster moving than
the radio lobe, then the kinematic ages of many CSOs would be
underestimated.  However, with such small sources, it is possible that
a true classical radio lobe has not yet been created and the hotspots
are the only indicator of the current phase of activity in CSOs.

\acknowledgments 

NEG gratefully acknowledges support from the NRAO Graduate Summer
Student Research Assistantship.  ABP thanks NRAO for travel support
and hospitality during part of this project.  This research has made
use of the NASA Astrophysics Data System.

\clearpage

%Figure 1 -- J0000
\begin{figure}
\figurenum{1}
\vspace{20.4cm}
\includegraphics{f1a.eps}
\includegraphics{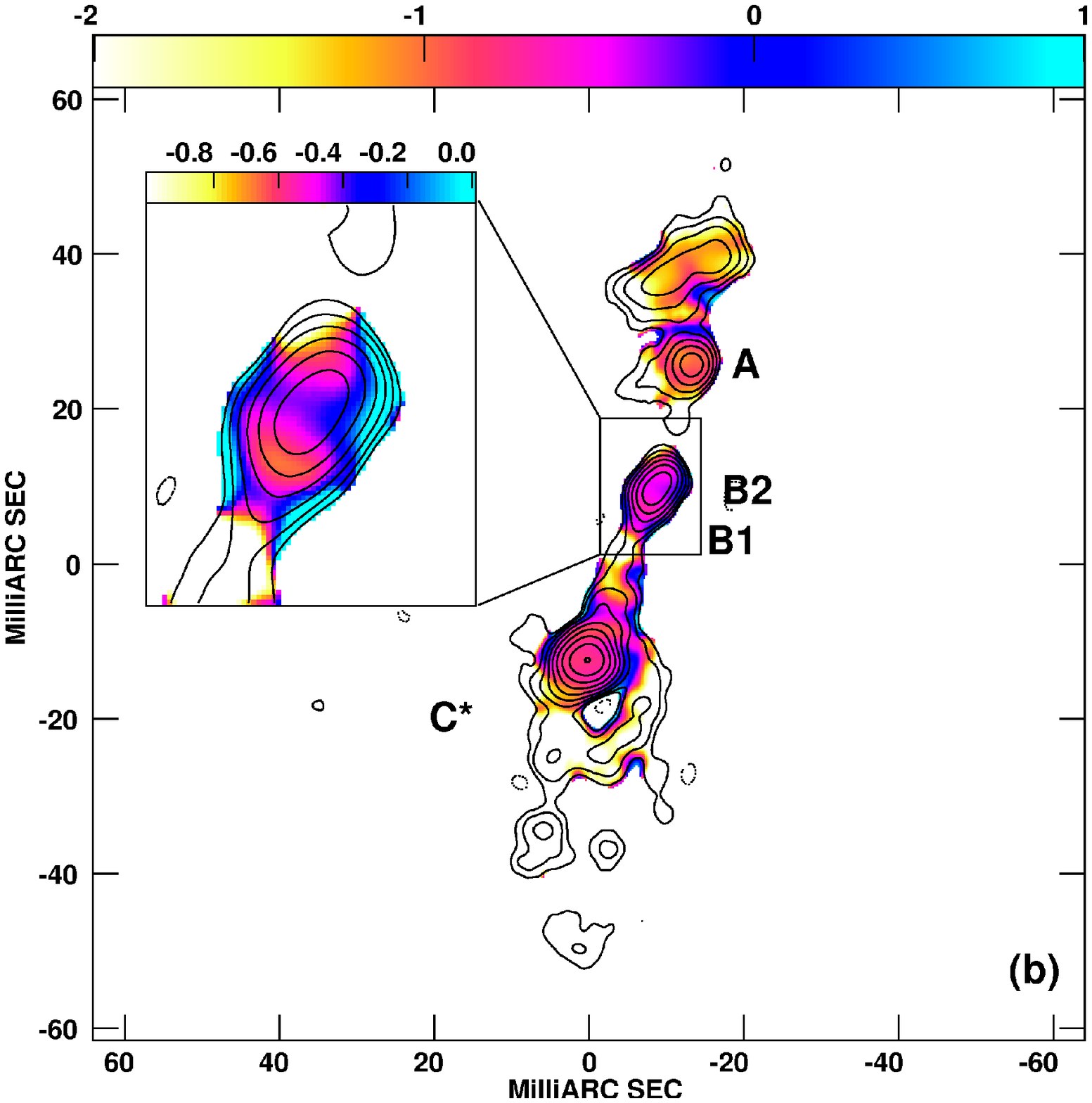}
\vspace{-2cm}
\caption{Total intensity contours of J0000$+$4054.  (a)~Electric
  polarization vectors over 8.4~GHz contours.  A vector length of
  1~mas corresponds to a polarized flux density of 0.05 mJy beam$^{-1}$; the smallest ticks correspond to $\sim$0.6~mJy~beam$^{-1}$.
  Contour levels begin at 0.25 mJy beam$^{-1}$ and increase by factors
  of 2.  (b)~Spectral index map over 4.8~GHz contours where $S_{\nu} \propto
  \nu^{\alpha}$.  Contour levels begin at 0.7~mJy~beam$^{-1}$ and
  increase by factors of 2.  The inset rescales the spectral indices
  for the B1 and B2 components.  A star indicates the reference
  component for motions.} 
\end{figure}
\clearpage

%Figure 2 -- J1826
\begin{figure}
\figurenum{2}
\vspace{20.4cm}
\includegraphics{f2a.eps}
\includegraphics{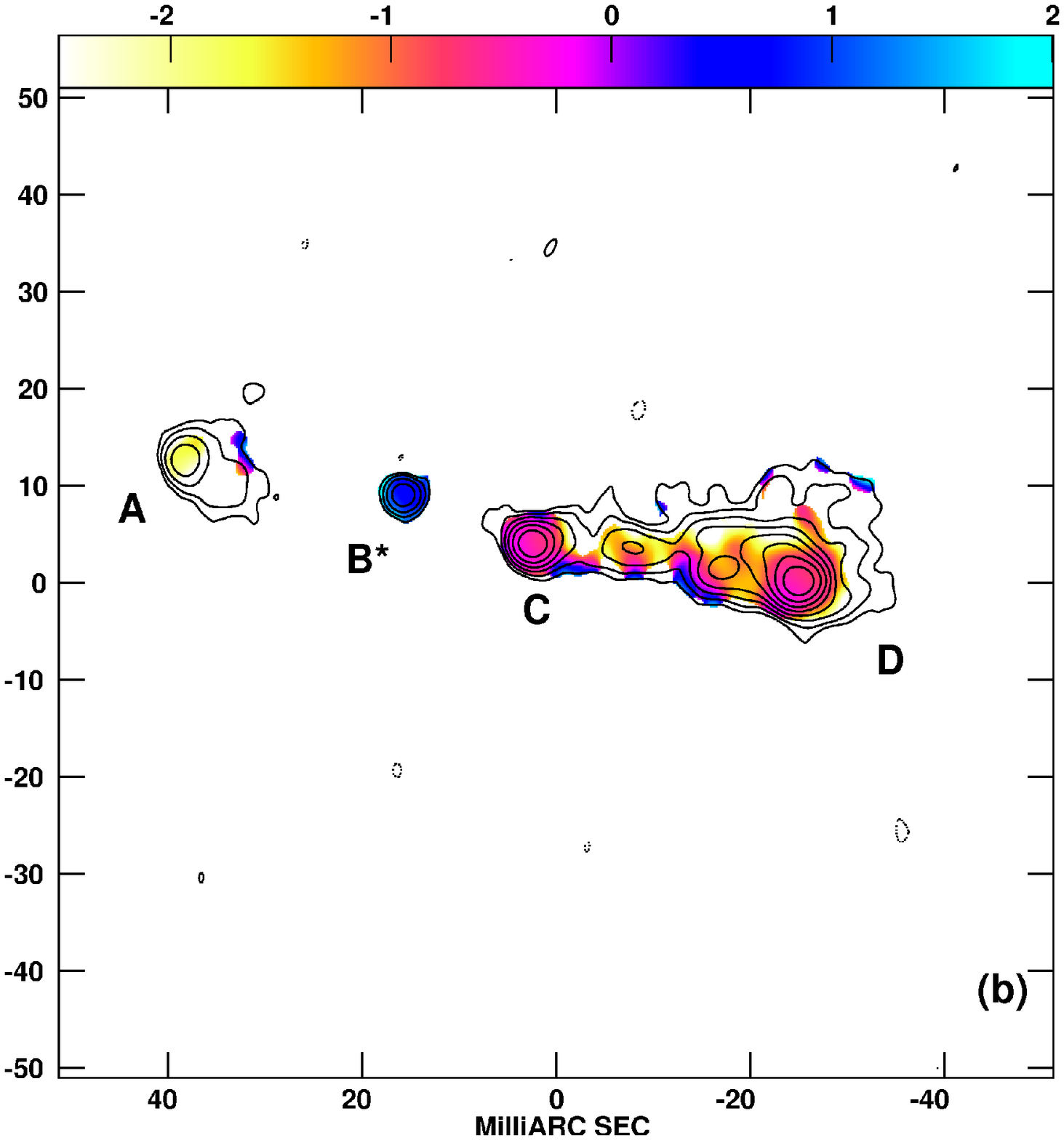}
\vspace{-2cm}
\caption{Total intensity contours of J1826$+$1831.  (a)~Electric
  polarization vectors over 8.4~GHz contours.  A vector length of
  1~mas corresponds to a polarized flux density of 0.10 mJy
  beam$^{-1}$; the smallest ticks correspond to $\sim$0.8~mJy~beam$^{-1}$.
  Contour levels begin at 0.25 mJy beam$^{-1}$ and increase by factors
  of 2.  (b)~Spectral index map over 4.8~GHz contours where $S_{\nu}
  \propto
  \nu^{\alpha}$.  Contour levels begin at 0.6~mJy~beam$^{-1}$ and
  increase by factors of 2.  A star indicates the reference
  component for motions.}
\end{figure}
\clearpage

%Figure 3 -- J1826 RM plot
\begin{figure}
\figurenum{3}
\vspace{20.4cm}
\includegraphics{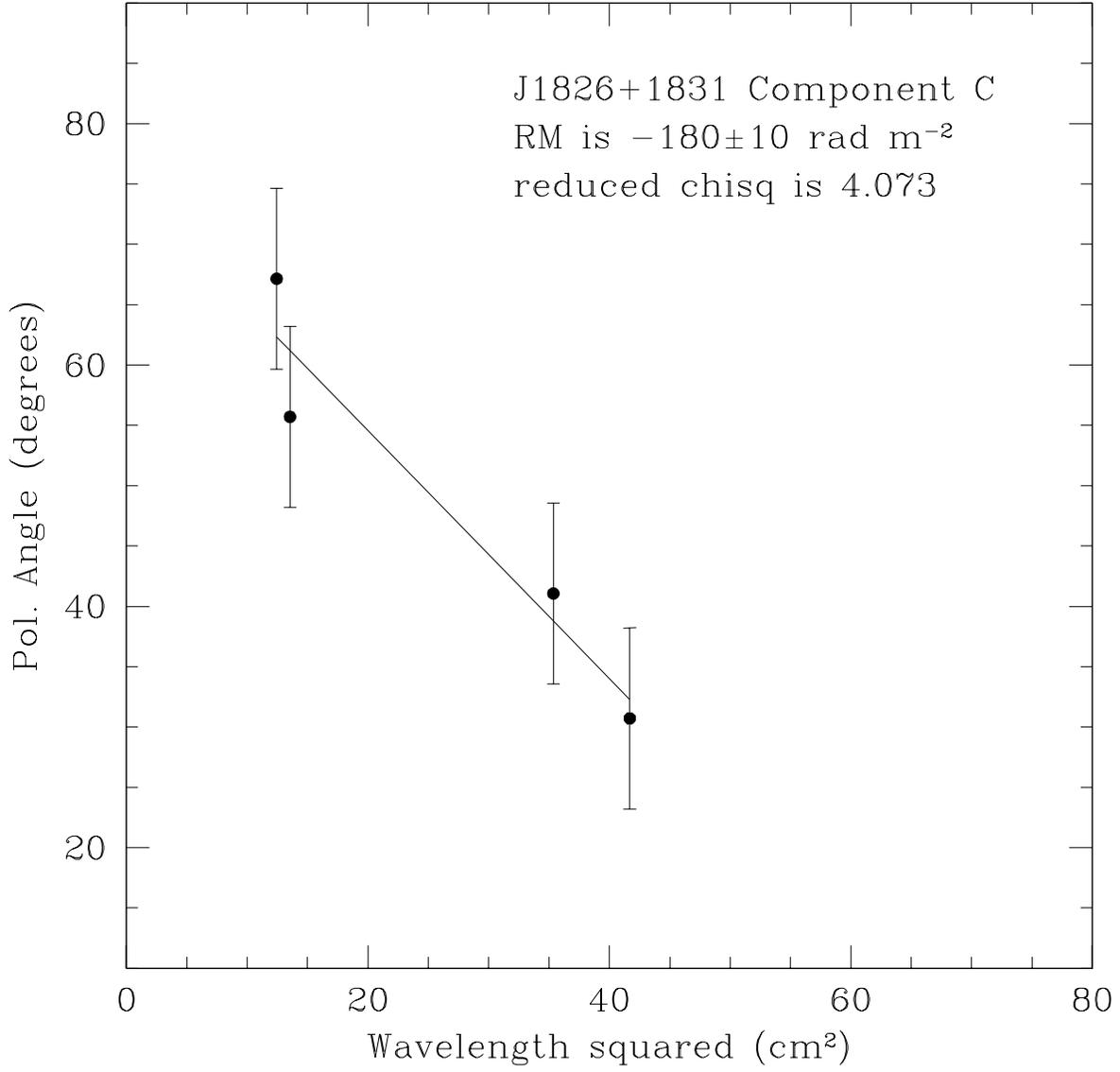}
\caption{Plot of polarization angle versus wavelength squared for
  component C of J1826$+$1831 where the slope of the least squares
  line is the rotation measure.}
\end{figure}
\clearpage

%Figure 4 -- J1826 motion plots
\begin{figure}
\figurenum{4}
\vspace{20.4cm}
\includegraphics{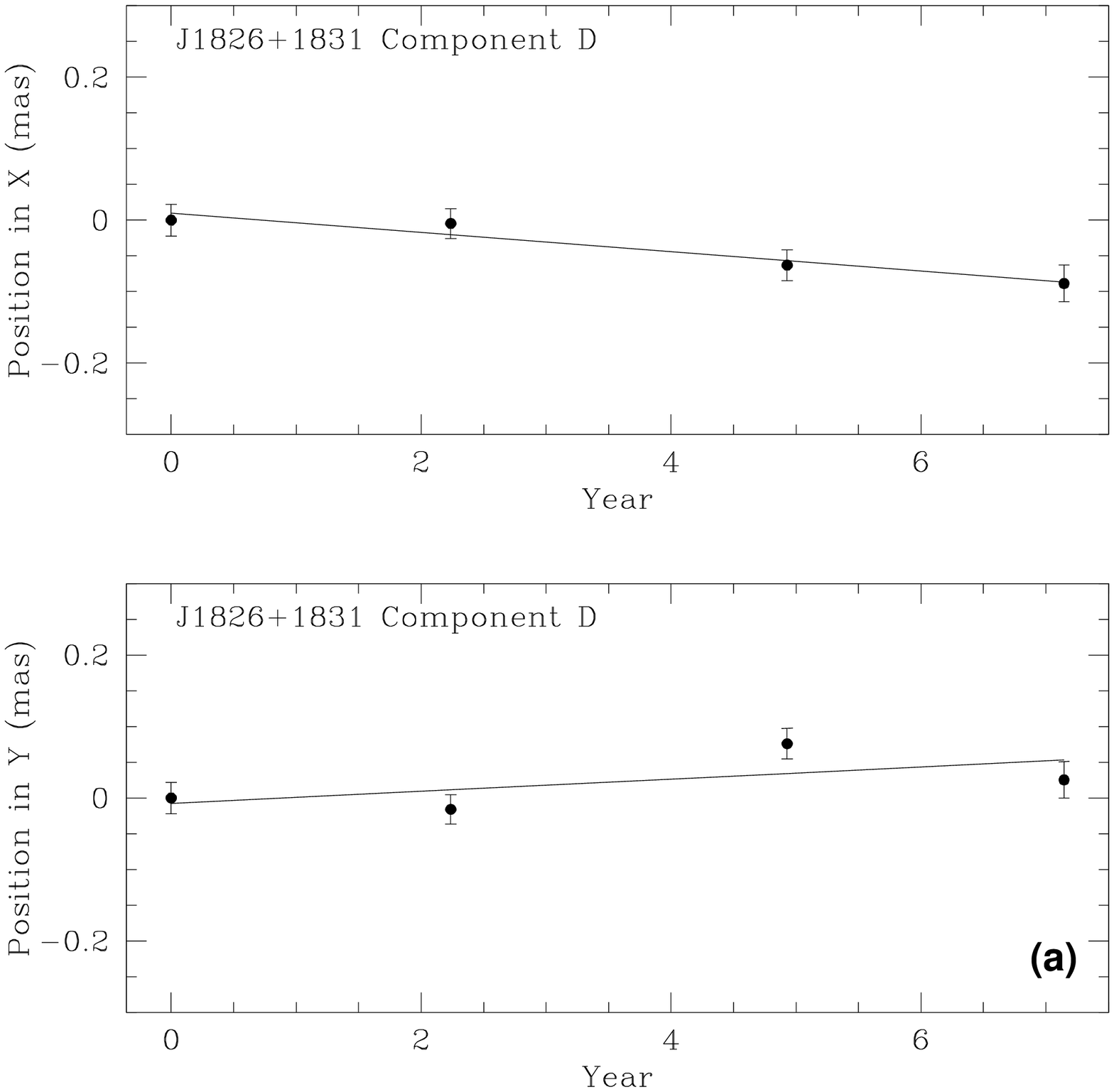}
\includegraphics{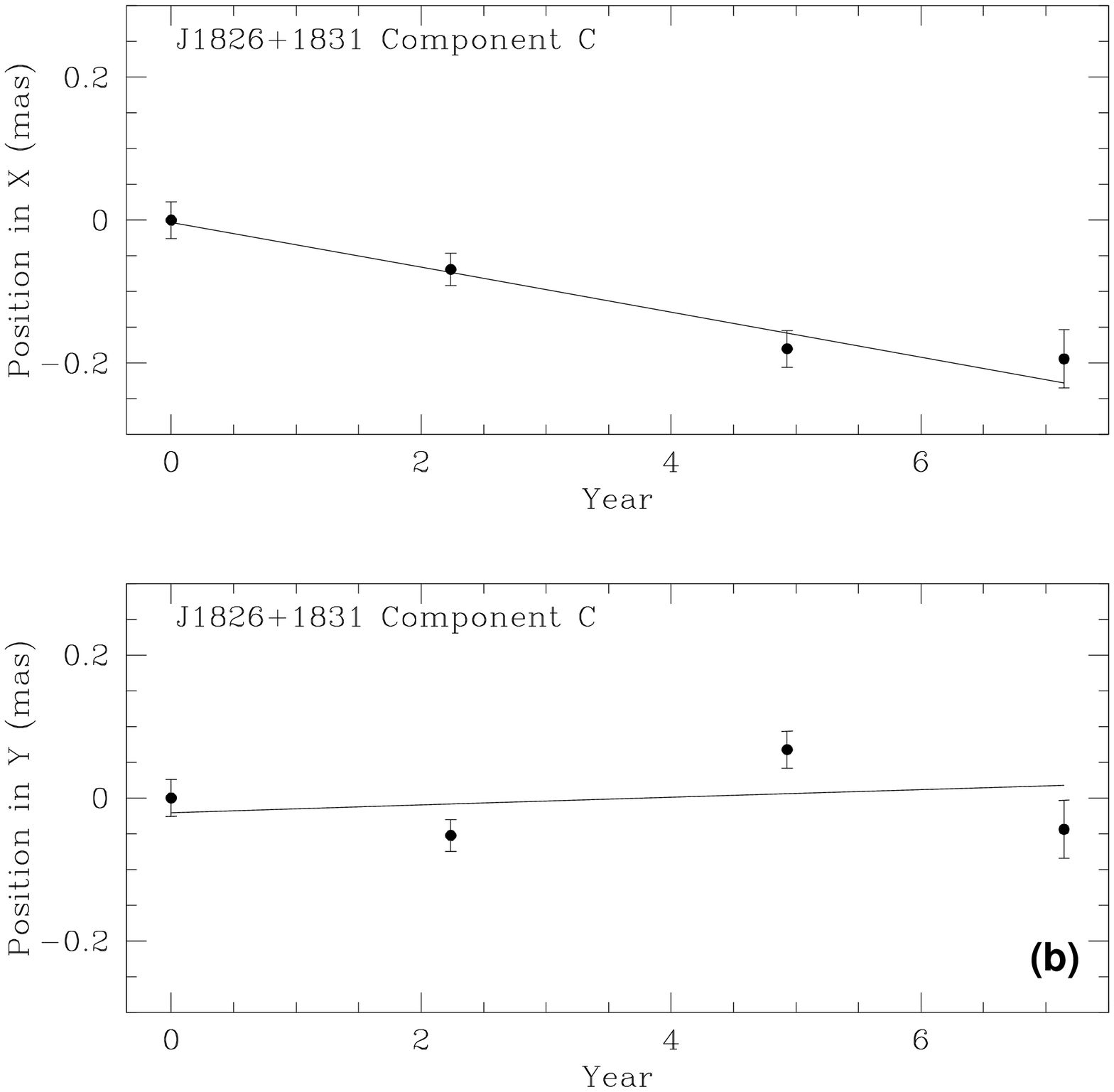}
\vspace{-1cm}
\caption{Proper motion plots of
  J1826$+$1831 with four epochs of 8.4~GHz models to the visibility
  data.  (a)~The slope for component D along the $x$ axis is
  $-$0.014$\pm$0.002~mas~yr$^{-1}$ with a reduced chi squared of 0.912.
  The slope along the $y$ axis is 0.008$\pm$0.004~mas~yr$^{-1}$ with a reduced chi
  squared of 7.487.
  (b)~The slope for component C along the $x$ axis is
  $-$0.031$\pm$0.003~mas~yr$^{-1}$ with a reduced chi squared of 2.385.
  The slope along the $y$ axis is 0.005$\pm$0.008~mas~yr$^{-1}$ with a reduced chi
  squared of 17.193.}
\end{figure}
\clearpage

%Figure 5 -- J1915 POL
\begin{figure}
\figurenum{5}
\vspace{20.4cm}
\includegraphics{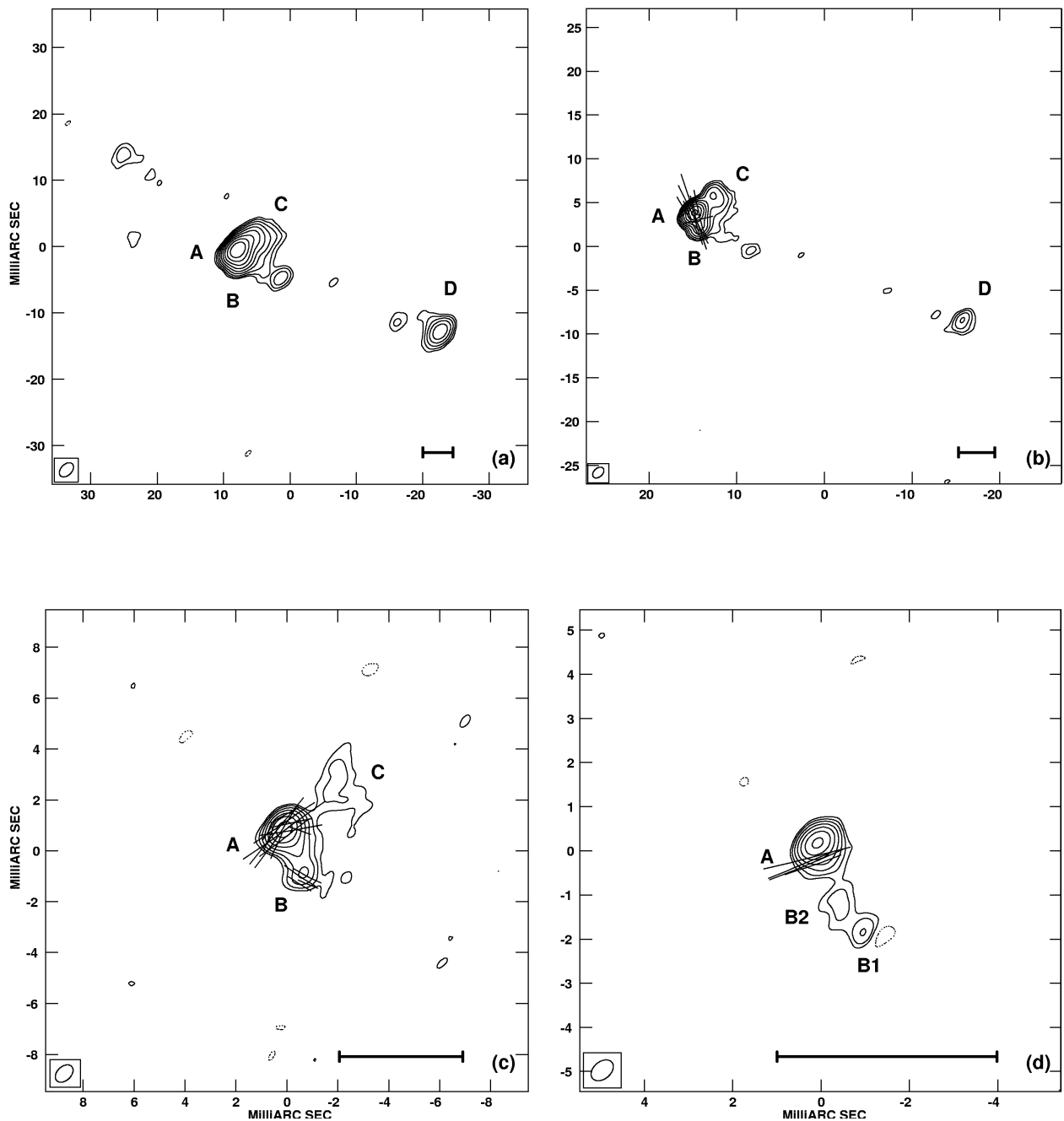}
\vspace{-1.5cm}
\caption{Total intensity plots of J1915$+$6548 at (a)~4.8~GHz,
  (b)~8.4~GHz, (c)~15.1~GHz, and (d)~22.2~GHz with electric
  polarization vectors overlayed.  Contour levels begin at
  0.4~mJy~beam$^{-1}$ in (a) and (b) and at 5.5~mJy~beam$^{-1}$ in (c)
  and (d) and increase by factors of 2.  A vector length of 1~mas
  corresponds to a polarized flux density of (b) 0.42~mJy~beam$^{-1}$,
  (c) 0.83~mJy~beam$^{-1}$, and (d) 1.7~mJy~beam$^{-1}$.  The bar in
  the lower right represents 5~mas or $\sim$45~pc.}
\end{figure}
\clearpage

%Figure 6 -- J1915 spectra
\begin{figure}
\figurenum{6}
\vspace{20.4cm}
\includegraphics{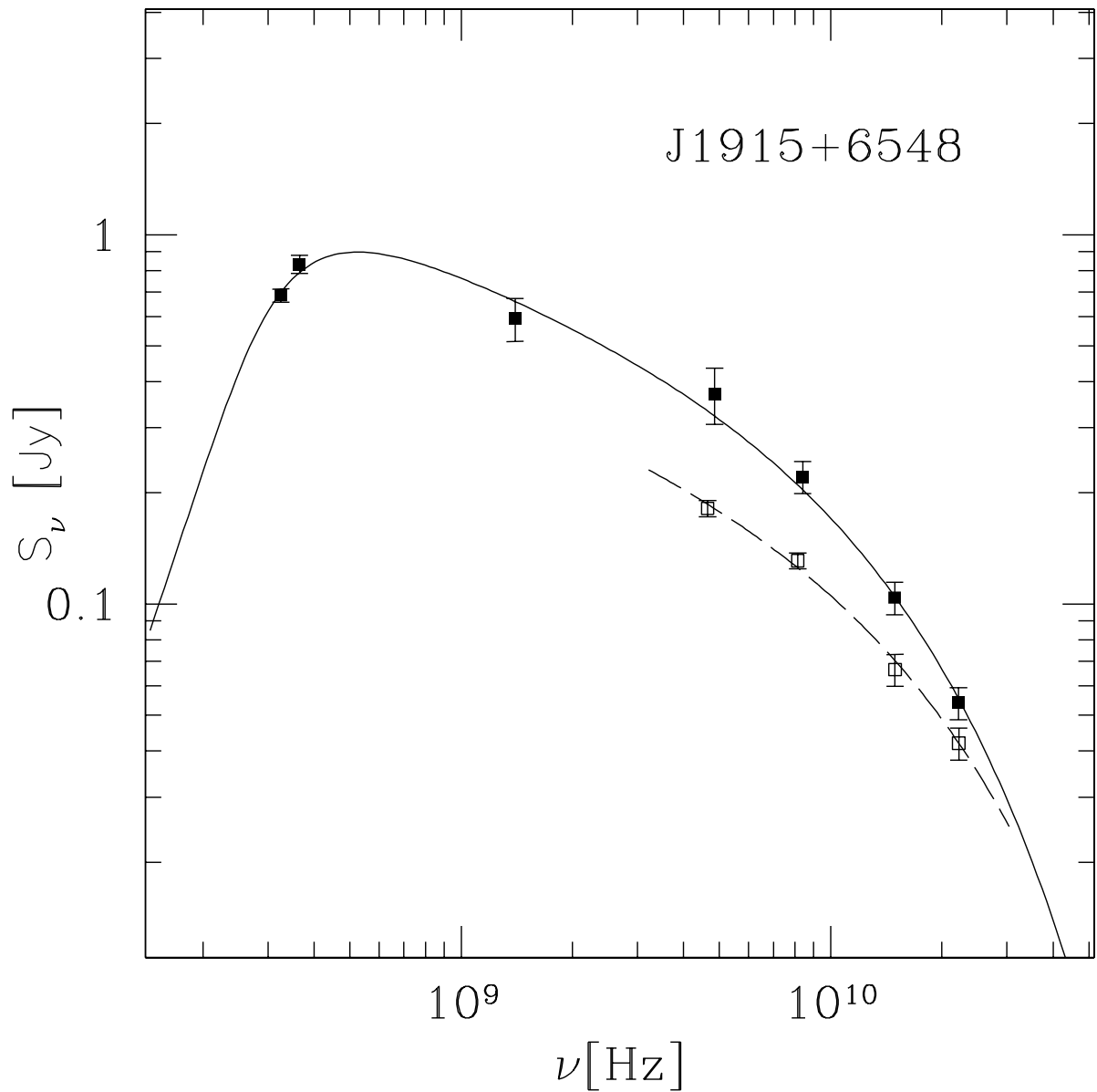}
\caption{Total power spectrum of J1915$+$6548 using data points from
the NASA Extragalactic Database and the data presented in this paper
(filled squares).  The break frequency is 17~GHz.  Overlayed is the
spectrum of component A which has a spectral index of $-$0.9 (open
squares).}
\end{figure}
\clearpage

%Figure 7 -- J1915 RM plot
\begin{figure}
\figurenum{7}
\vspace{20.4cm}
\includegraphics{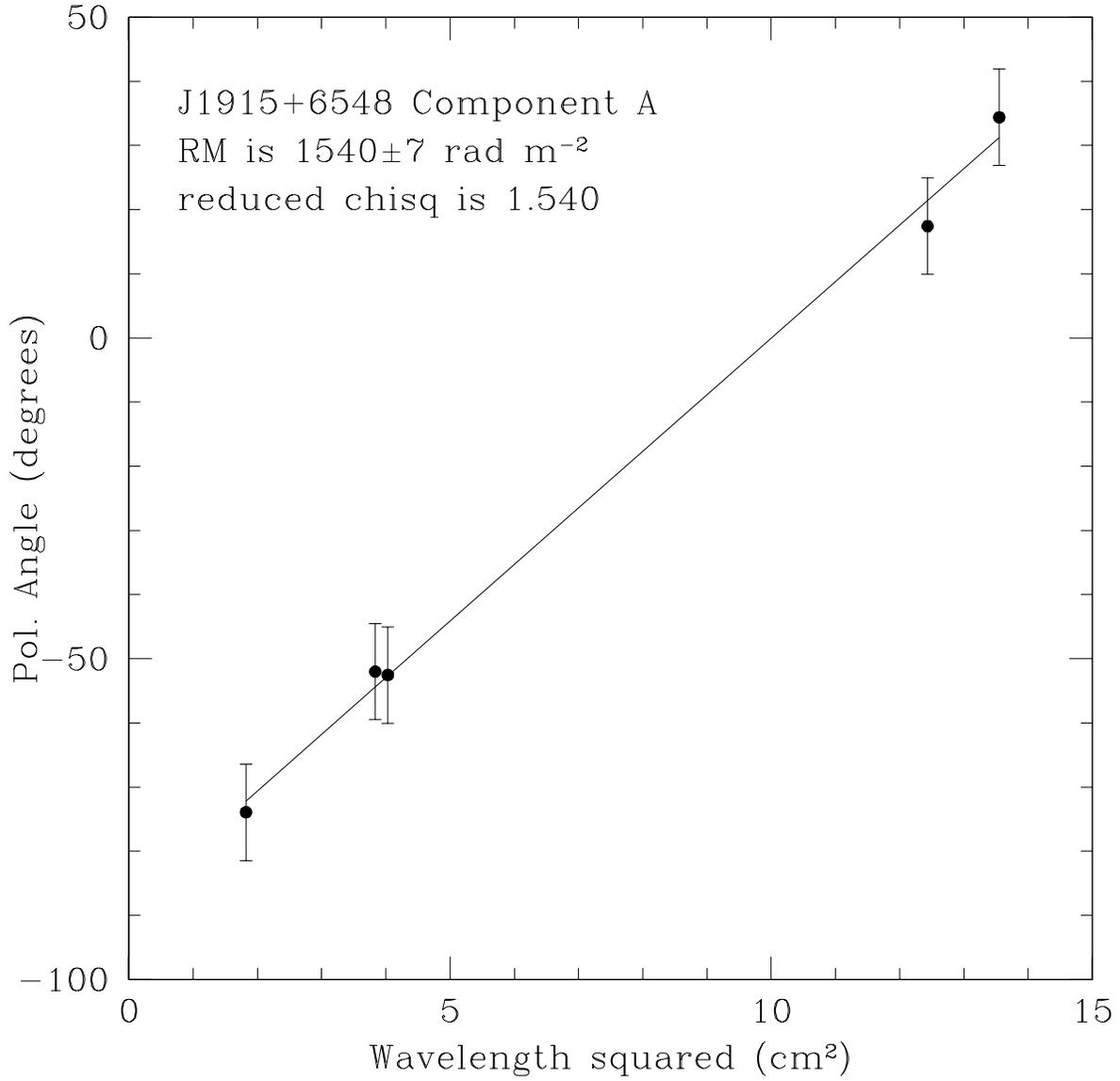}
\caption{Plot of polarization angle versus wavelength squared for
  component A of J1915$+$6548 where the slope of the least squares
  line is the rotation measure.  Note that $x$ and $y$ scales are
  different. }
\end{figure}
\clearpage

%Figure 8 -- J1915 motions
\begin{figure}
\figurenum{8}
\vspace{20.4cm}
\includegraphics{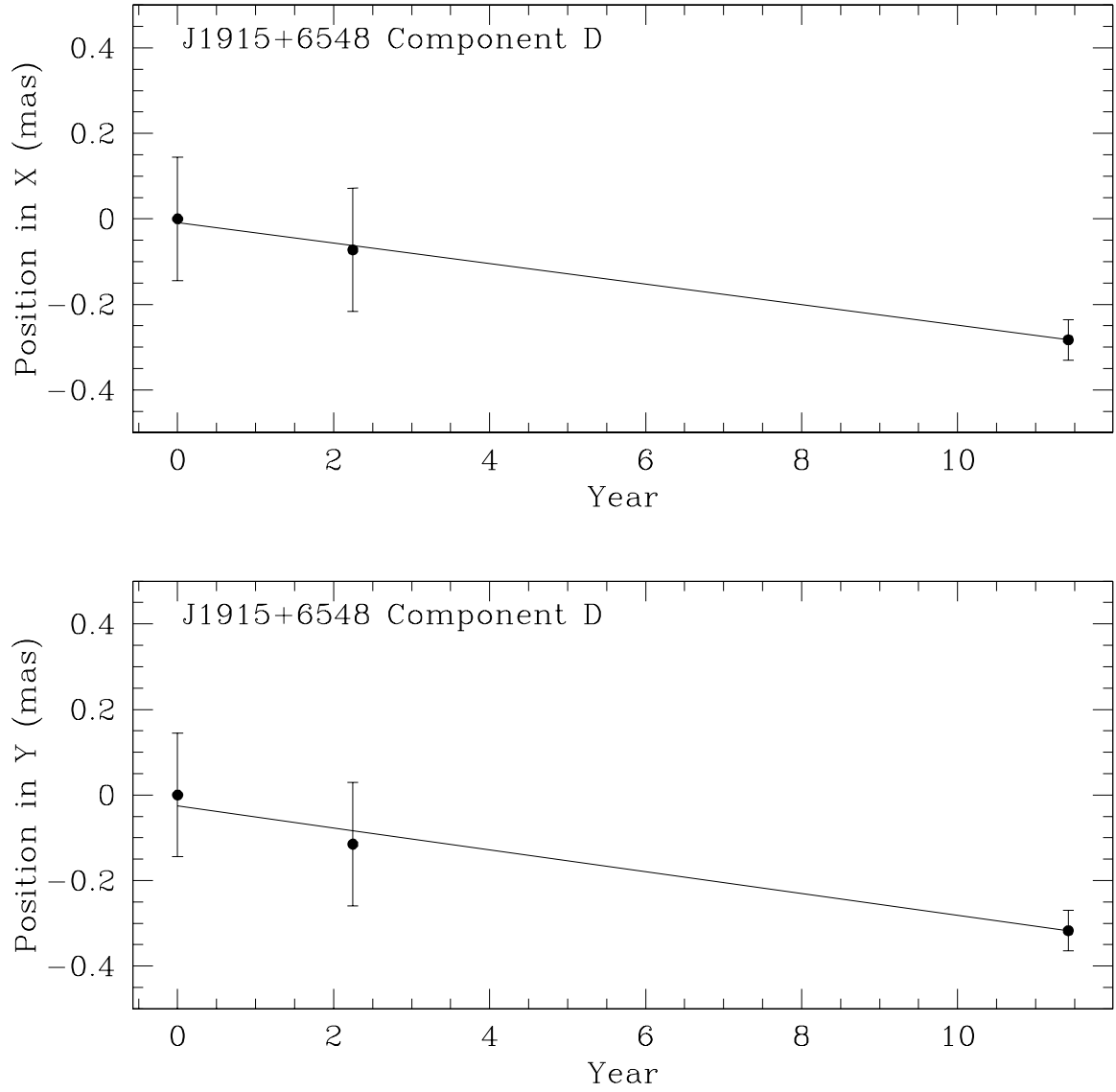}
\caption{Proper motion plots of J1915$+$6548 with three epochs of
  model fit to the 4.8~GHz visibility data.  The slope for component D
  along the $x$ axis is $-$0.024$\pm$0.002~mas~yr$^{-1}$ with a
  reduced chi squared of 0.118 The slope along the $y$ axis is
  $-$0.026$\pm$0.005~mas~yr$^{-1}$ with a reduced chi squared of
  1.153.}
\end{figure}
\clearpage

%Figure 9 -- Beta Theta, vsep
\begin{figure}
\figurenum{9}
\vspace{20.4cm}
\includegraphics{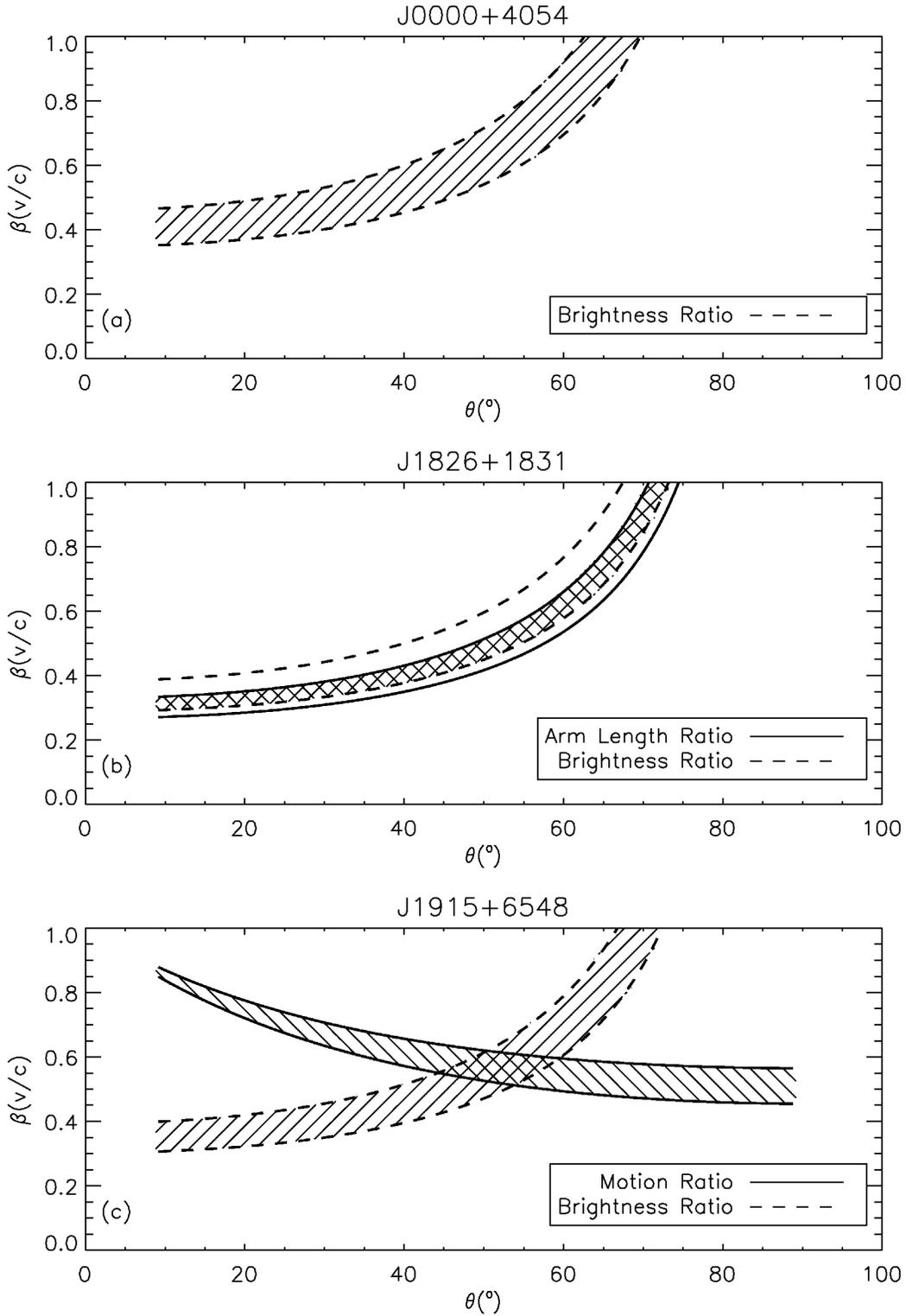}
\vspace{-0.5cm}
\caption{Plots of loci for intrinsic jet velocity, $\beta$,  vs. angle to the line of sight, $\theta$, for (a)~J0000$+$4054,
(b)~J1826$+$1831, and (c)~J1915$+$46548.  The brightness ratios are
calculated from two different models with the hatching to indicate that
the values lie between the lines.  The arm length and motion ratios
are hatched in between the upper and lower limits due to experimental
error.  Cross-hatching indicates where the regions overlap.
Single-hatching is removed from (b) for clarity.}

\end{figure}
\clearpage

\begin{deluxetable}{lccccc}
\tabletypesize{\footnotesize}
\tablewidth{0pt}
\tablecolumns{6}
\tablecaption{Observational Parameters\label{tab1}}
\tablehead{\colhead{Source} & \colhead{Date} & \colhead{Freq.} &
  \colhead{Time} & \colhead{Bandwidth} & \colhead{rms}\\ 
  \colhead{} & \colhead{} & \colhead {(GHz)} & \colhead{(min)} &
  \colhead{(MHz)} & \colhead{(mJy)}\\
  \colhead{(1)} & \colhead{(2)} & \colhead{(3)} & \colhead{(4)} &
  \colhead{(5)} & \colhead{(6)}}
\startdata
J0000$+$4054 & 20050212 & 4.6 & 200 & 16 & 0.18 \\
 & 20050212 & 5.0 & 200 & 16 & 0.12\\
 & 20050212 & 8.2 & 200 & 16 & 0.12\\
 & 20050212 & 8.5 & 200 & 16 & 0.09\\
J1826$+$1831 & 20050218 & 4.6 & 212 & 16 & 0.17\\
 & 20050218 & 5.0 & 212 & 16 & 0.18\\
 & 20050218 & 8.2 & 212 & 16 & 0.20\\
 & 20050218 & 8.5 & 212 & 16 & 0.19\\
J1915$+$6548 & 20041111 & 4.6 & 58 & 16 & 0.23\\
 & 20041111 & 5.0 & 58 & 16 & 0.18\\
 & 20041111 & 8.2 & 39 & 16 & 0.17\\
 & 20041111 & 8.5 & 39 & 16 & 0.15\\
 & 20041111 & 14.9 & 97 & 16 & 0.26\\
 & 20041111 & 15.3 & 97 & 16 & 0.21\\
 & 20041111 & 22.2 & 135 & 32 & 0.18\\
\enddata
\tablenotetext{*}{
Notes - (1) J2000 source name; (2) Date of observation; (3) Frequency
in GHz; (4) Integration time in minutes; (5) Bandwidth in MHz; (6) rms
noise in mJy.  Note that the rms for each fully averaged frequency, 4.8, 8.4,
and 15.1~GHz, is generally a factor of $\sqrt{2}$ lower.}
\end{deluxetable}
\clearpage

\begin{deluxetable}{lllccccccc}
\tabletypesize{\footnotesize}
\tablewidth{0pt}
\tablecolumns{10}
\tablecaption{Source Parameters\label{tab2}}
\tablehead{\colhead{Name} & \colhead{RA} &
  \colhead{Dec} &\colhead{ID} & \colhead{$M_v$} & \colhead{$z$} &
  \colhead{$S_{\mbox{5GHz}}$} & \colhead{$S_{\mbox{8GHz}}$}  &
  \colhead{$S_{\mbox{15GHz}}$} & \colhead{$S_{\mbox{22GHz}}$} \\
  \colhead{(1)} & \colhead{(2)} & \colhead{(3)} & \colhead{(4)} &
  \colhead{(5)} & \colhead{(6)} & \colhead{(7)} & \colhead{(8)} &
  \colhead{(9)} & \colhead{(10)}}
\startdata
J0000$+$4054 & 00 00 53.081551 & $+$40 54 01.79335  &
G & 21.4 & ...& 521 & 322 & ... & ... \\
J1826$+$1831 & 18 26 17.710882 & $+$18 31 52.88973 & ... &
... & ...& 427 & 279 & ... & ... \\
J1915$+$6548 & 19 15 23.819114 & $+$65 48 46.38505 & G & 18.2 & 0.486
& 331 & 202 & 104 & 54 \\
\enddata
\tablenotetext{*}{
Notes - (1) J2000 source name; (2) Right
ascension and (3) Declination in J2000 coordinates from the VLBA
Calibrator Survey by Beasley et al. 2002; (4)
Optical host galaxy identification; (5) Optical magnitude; (6)
Redshift; (7) Total
flux density at 4.8~GHz in mJy; (8) Total flux density at 8.4~GHz in
mJy; (9) Total flux density at 15.1~GHz in mJy; (10) Total flux
density at 22.2~GHz in mJy.}
\end{deluxetable}
\clearpage

\begin{deluxetable}{lccccccccc}
\tabletypesize{\footnotesize}
\tablewidth{0pt}
\tablecolumns{10}
\tablecaption{CSO Model Parameters\label{tab3}}
\tablehead{\colhead{} & \colhead{} & \colhead{$b_{maj}$} &
  \colhead{$b_{min}$} & \colhead{$\phi$} & \colhead{$S$} &
  \colhead{$P$} & \colhead{$\mu$} & \colhead{$v$} & \colhead{Kinetic Age} \\
  \colhead{Source} & \colhead{Component} & \colhead{(mas)} &
  \colhead{(mas)} & \colhead{(deg.)} & \colhead{(mJy)} &
  \colhead{(mJy)} & \colhead{(mas yr$^{-1}$)} & \colhead{($c$)} & \colhead{(yr)} \\
  \colhead{(1)} & \colhead{(2)} & \colhead{(3)} & \colhead{(4)} &
  \colhead{(5)} & \colhead{(6)} &\colhead{(7)} &  \colhead{(8)} &
  \colhead{(9)} & \colhead{(10)}}
\startdata
J0000$+$4054 & A & 2.28 & 1.71 & $-$3.4 & 14 & $<$0.2 & $<$0.066 & ... & 
$>$610\\
 & B1 & 0.91 & 0.42 & $-$83.6 & 19 & $<$0.2 & ... & ... & ... \\
 & B2 & 0.63 & 0.63 & ... & 28 & $<$0.2 & ... & ... & ... \\
 & C & 3.01 & 2.41 & $-$65.4 & 205 & 2.1 & Reference & ... & ... \\
J1826$+$1831 & A & 8.05 & 4.51 & 24.6 & 12 & $<$0.5 & ... & ... & ... \\
 & B & 0.45 & 0.45 & ... & 11$^a$ & $<$0.5 & Reference & ... & ... \\
 & C & 0.96 & 0.77 & 43.3 & 32 & 2.3 & 0.032$\pm$0.003 & ... & 450$\pm$43$^b$ \\
 & D & 1.74 & 1.11 & $-$71.3 & 115 & $<$0.5 & 0.015$\pm$0.003 & ... & 
2600$\pm$490 \\
J1915$+$6548 & A & 0.46 & 0.32 & 25.2 & 183 & $<$0.3 & Reference &
... & ... \\
 & B & 1.88 & 0.86 & 29.8 & 72 & $<$0.3 & ... & ... & ... \\
 & C & 1.65 & 1.37 & 24.9 & 36 & $<$0.3 & ... & ... & ... \\
 & D & 1.30 & 0.88 & 17.6 & 15 & $<$0.3 & 0.035$\pm$0.004 &
1.02$\pm$0.11  & 940$\pm$110 \\ 
\enddata
\tablenotetext{*}{
Notes - (1) J2000 source name; (2) Component name; (3) Major and (4)
minor axes of Gaussian model component; (5) Position angle of major axis;
(6) Integrated flux density of Gaussian model component; (7) Polarized intensity, or 3$\sigma$
limit; (8) Relative proper motion; (9) Relative proper motion in terms
of $c$ if $z$ is available; (10) Kinematic age estimate.  For
J0000$+$4054 and J1826$+$1831, fluxes are at 8.4~GHz at 2005.118.  For
J1915$+$6548, fluxes are at 4.8~GHz at 2004.863.}
\tablenotetext{a}{
Correction: Gugliucci et al. (2005) erroneously listed the flux
densities of component B as 80, 90 and 70 mJy in epochs 1, 2, and 3.
The correct fluxes are 8, 9, and 7 mJy, respectively.  This gives a core
fraction of 3\%.}
\tablenotetext{b}{This is not a source age, but a kinematic
  age estimate for a jet component.}
\end{deluxetable}
\clearpage

\begin{deluxetable}{lccccc}
\tabletypesize{\footnotesize}
\tablewidth{0pt}
\tablecolumns{6}
\tablecaption{Hotspot Ratios for Polarized and Unpolarized CSOs\label{tab4}}
\tablehead{\colhead{} & \colhead{} & \colhead{$S_{1}$} &
  \colhead{$S_{2}$} & \colhead{Ratio} & \colhead{Core} \\
  \colhead{} & \colhead{Source} & \colhead{(mJy)} & \colhead{(mJy)} &
  \colhead{($S_{2} \over S_{1}$)} & \colhead{Fraction} \\
  \colhead{(1)} & \colhead{(2)} & \colhead{(3)} & \colhead{(4)} &
  \colhead{(5)} & \colhead{(6)}}
\startdata
Polarized Sources & J0000$+$4054 & 205 & 14 & 0.07 & $<$0.0006 \\
 & J1826$+$1831 & 115 & 12 & 0.10 & 0.04 \\
 & J1915$+$6548 & 153 & 6 & 0.04 & $<$0.001 \\
 & & & & \\
Unpolarized Sources & J0003$+$4807 & 38 & 13 & 0.34 & 0.04 \\
 & J0204$+$0903 & 120 & 68 & 0.57 & 0.20 \\
 & J0427$+$4133 & 65 & 28 & 0.43 & 0.86 \\
 & J0620$+$2102 & 156 & 88 & 0.56 & $<$0.001 \\
 & J0754$+$5324 & 39 & 36 & 0.92 & $<$0.003 \\
 & J1111$+$1955 & 98 & 76 & 0.78 & $<$0.002 \\
 & J1143$+$1834 & 130 & 99 & 0.76 & $<$0.001 \\
 & J1414$+$4554 & 64 & 51 & 0.80 & $<$0.003 \\
 & J1546$+$0026 & 213 & 110 & 0.52 & 0.40 \\
 & J1734$+$0926 & 183 & 119 & 0.65 & $<$0.0006 \\
 & J1816$+$3457 & 135 & 59 & 0.44 & $<$0.001 \\
 & J2203$+$1007 & 145 & 53 & 0.37 & $<$0.001 \\
\enddata
\tablenotetext{*}{
Notes - (1) Category; (2) J2000 source name; (3) Integrated flux
density of Gaussian model component of brighter hotspot at 8.4~GHz;
(4) Integrated flux density of Gaussian model component of dimmer
hotspot at 8.4~GHz; (5) Ratio of dimmer hotspot as
compared to brighter hotspot; (6) Fraction of total flux that is
attributed to the core (using 3$\sigma$ limit for core flux if not
detected).  Fluxes for unpolarized CSOs from
2000.227 in GTPG.  Fluxes for polarized CSOs and candidate from this
paper.}
\end{deluxetable}
\clearpage

\end{document}